\documentclass{article}
\usepackage{amsmath}
\usepackage[psamsfonts]{amssymb}
\usepackage{cmmib57}
\usepackage{diagrams}
\newcommand{\bPf}{\par\vspace*{-4pt}\indent{\sc Proof.}\enskip}
\newcommand{\ePf}{\medskip}
\def\QED{\hskip0.1em\hfill\null\ \null\nobreak\hfill\kern3pt\vbox{\hrule\hbox
   {\vrule\kern1pt\vbox{\kern1.7pt\hbox{$\scriptscriptstyle{QED}$}
    \kern0.2pt}\kern1pt\vrule}\hrule}}

\def\END{\hskip0.1em\hfill\null\ \null\nobreak\hfill\kern3pt\vbox{\hrule\hbox
   {\vrule\kern1pt\vbox{\kern1.7pt\hbox{$\,\,\,\vspace{5pt}$}
    \kern0.2pt}\kern1pt\vrule}\hrule}}
\newtheorem{theorem}{Theorem}
\newtheorem{lemma}{Lemma}
\newtheorem{corollary}{Corollary}
\newtheorem{proposition}{Proposition}
\newtheorem{remark}{Remark}
\newtheorem{definition}{Definition}
\newtheorem{example}{Example}
\newcommand{\bCd}{\bEq\begin{CD}}
\newcommand{\eCd}{\end{CD}\eEq}
\newcommand{\bcd}{\beq\begin{CD}}
\newcommand{\ecd}{\end{CD}\eeq}
\newcommand{\ben}{\begin{enumerate}}
\newcommand{\een}{\end{enumerate}}
\newcommand{\bEq}{\begin{eqnarray}}
\newcommand{\eEq}{\end{eqnarray}}
\newcommand{\beq}{\begin{eqnarray*}}
\newcommand{\eeq}{\end{eqnarray*}}
\newcommand{\bDf}{\begin{definition}\em}
\newcommand{\eDf}{\end{definition}}
\newcommand{\bLm}{\begin{lemma}}
\newcommand{\eLm}{\end{lemma}}
\newcommand{\bPr}{\begin{proposition}}
\newcommand{\ePr}{\end{proposition}}
\newcommand{\bTh}{\begin{theorem}}
\newcommand{\eTh}{\end{theorem}}
\newcommand{\bCr}{\begin{corollary}}
\newcommand{\eCr}{\end{corollary}}
\newcommand{\bRm}{\begin{remark}\em}
\newcommand{\eRm}{\end{remark}}
\newcommand{\bEx}{\begin{example}\em}
\newcommand{\eEx}{\end{example}}
\newcommand{\ie}{{\em i.e$.$} }
\newcommand{\eg}{{\em e.g$.$} }
\newcommand{\der}{\partial}

\DeclareMathOperator{\byd}{{\raisebox{.1ex}{:}{=}}}
\newcommand{\ucar}[1]{\underset{#1}{\times}}

\newcommand{\udir}[1]{\underset{#1}{\oplus}}

\newcommand{\uten}[1]{\underset{#1}{\otimes}}

\newcommand{\balp}{\boldsymbol{\alp}}

\newcommand{\cD}{\mathcal{D}}
\newcommand{\cE}{\mathcal{E}}

\newcommand{\cH}{\mathcal{H}}

\newcommand{\bU}{\boldsymbol{U}}

\newcommand{\bX}{\boldsymbol{X}}
\newcommand{\bY}{\boldsymbol{Y}}

\newcommand{\sub}{\subset}

\newcommand{\wed}{\wedge}
\newcommand{\com}{\!\circ\!}
\newcommand{\ten}{\!\otimes\!}

\newcommand{\alp}{\alpha}

\newcommand{\gam}{\gamma}

\newcommand{\lam}{\lambda}
\newcommand{\sig}{\sigma}

\newcommand{\ome}{\omega}
\newcommand{\Gam}{\Gamma}
\newcommand{\Del}{\Delta}

\newcommand{\Ome}{\Omega}

\newcommand{\vartht}{\vartheta}

\newcommand{\Tht}{\Theta}
\newcommand{\For}{{\Lambda}}

\title{{\bf{\large A New Geometric Proposal for the Hamiltonian \\
Description of Classical
Field Theories\,\thanks{The paper is in final form and will not be published elsewhere.}}}}
\bigskip
\author{\large{Mauro Francaviglia,\, Marcella Palese\thanks{
Both supported by GNFM of INdAM, MURST and University of
Torino.}}
\\
\large{and}
\\
\large{Ekkehart Winterroth\thanks{Supported by GNFM of INdAM and University
of Erlangen--N\"urnberg.}}
\\
{\footnotesize Department of Mathematics, University of Torino}
\\{\footnotesize Via C. Alberto 10, 10123 Torino, Italy}
\\{\footnotesize e--mails: {\sc francaviglia@dm.unito.it, palese@dm.unito.it,}}
\\{\footnotesize {\sc ekkehart@dm.unito.it}}
}
\date{}
\begin{document}

\maketitle

\begin{abstract}

We consider the geometric formulation of the Hamiltonian formalism for field theory in
terms of {\em Hamiltonian connections} and {\em multisymplectic forms}. In this
framework the covariant Hamilton equations for Mechanics and field theory are defined
in terms of multisymplectic $(n+2)$--forms, where $n$ is the dimension of the basis
manifold, together with connections on the configuration bundle. We provide a new geometric
Hamiltonian description of field theory, based on the introduction of a suitable {\em
composite fibered bundle} which plays the role of an {\em extended configuration
bundle}. Instead of fibrations over an
$n$--dimensional base manifold $\bX$, we consider {\em fibrations over a line bundle $\Tht$
fibered over $\bX$}. 
The concepts of {\em extended
Legendre bundle}, {\em Hamiltonian connection}, {\em Hamiltonian form} 
and {\em covariant Hamilton equations} are introduced and put in relation with the corresponding
standard concepts in the polymomentum approach to field theory. 

\medskip



\end{abstract}

\newpage

%------------------------------------------------------------------------------------------
\section{Introduction}\label{1}
%------------------------------------------------------------------------------------------

A geometric formulation of the Hamiltonian formalism for field theory in
terms of {\em Hamiltonian connections} and {\em multisymplectic forms} was developed in
\cite{MaSa00,Sar95,Sar98}. We recall that, in this framework, the covariant Hamilton
equations for Mechanics and field theory are defined in terms of multisymplectic $(n+2)$--forms,
where $n$ is the dimension of the basis manifold, together with connections on the 
configuration bundle.

We provide here a new geometric Hamiltonian description of field theory, based on
the introduction of a suitable {\em composite fibered bundle} which plays the role of
an {\em extended configuration bundle}. One of the main features of this approach
is that one can describe the polymomenta and other objects appearing in the {\em polymomentum}
formulation of field theory (see \eg
\cite{Ded77,Go91,Kan98,Kij73,KiTu79,Krup01} and references therein) in terms of
differential forms with values in the vertical tangent bundle of an appropriate line
bundle $\Tht$. The introduction of the line bundle $\Tht$ can be here understood as a suitable way
of describing the {\em gauge} character appearing in the Hamiltonian formalism for field
theory (see \cite{Kij74} for a nice introduction to this topic). Instead of
bundles over an
$n$--dimensional base manifold $\bX$, we consider {\em fibrations over a line bundle $\Tht$
fibered over $\bX$}; 
the concepts of {\em event bundle}, {\em configuration bundle} and {\em Legendre bundle} are then
introduced following the analogous setting introduced in \cite{MaSa00,Sar95,Sar98} for Mechanics and
for the polymomentum approach to field theory. Moreover, {\em Hamiltonian connections}, {\em
Hamiltonian forms} and {\em covariant Hamilton equations} can be suitably described in this
framework. This new approach takes into account the existence of more than one independent variable
in field theory, but enables us to keep as far as possible most of the nice features of
time--dependent Hamiltonian Mechanics. 

Already in the seventies, Kijowski
stressed the prominent role of the symplectic structures in field theories
\cite{Kij73,Kij74,KiSz76,KiTu79}. Our approach can provide a suitable geometric interpretation
of the canonical theory of gravity and gravitational energy, as presented in \cite{Kij01}, 
where the local line bundle coordinate
$\tau$ plays the role of a {\em parameter} and enables one to consider the gravitational energy
as a `{\em gravitational charge}'.

In Section \ref{2} we state the general framework of composite fiber bundles, their jet
prolongations and composite connections. Section \ref{3} contains the main result of this note, \ie
Theorem \ref{final}, which relates the {\em abstract Hamiltonian dynamics} introduced  here with the
standard Hamilton--De Donder equations (see \cite{Krup01} for a detailed review on the topic and
recent developments).
Furthermore, since it stresses
the underlying algebraic structure of field theory, this `extended' approach turns out to be very
promising, \eg {\em via} the application of some results concerned
with a new
$K$--theory for vector bundles carrying this special kind of multisymplectic structure (see
\cite{Win01}). However, this topic will be developed elsewhere.

%------------------------------------------------------------------------------------------
\section{Jets of fibered manifolds and connections}\label{2}
%------------------------------------------------------------------------------------------

The general framework is a fibered bundle $\pi : \bY \to \bX$,
with $\dim \bX = n$ and $\dim \bY = n+m$ 
and, for $r \geq 0$, its jet manifold $J_r\bY$. We recall the natural fiber 
bundles
$\pi^r_s : J_r\bY \to J_s\bY$, $r \geq s$, $\pi^r : J_r\bY \to \bX$, and, 
among these, the {\em affine\/} fiber bundles $\pi^r_{r-1}$.
We denote by $V\bY$ the vector subbundle of the tangent 
bundle $T\bY$ formed by vectors on $\bY$ which are vertical with respect 
to the fibering $\pi$ (see {\em e.g.} 
\cite{Sau89}).

Greek indices $\lam ,\mu ,\dots$ run from $1$ to $n$ and they label base 
coordinates, while
Latin indices $i,j,\dots$ run from $1$ to $m$ and label fibre coordinates,
unless otherwise specified.
We denote multi--indices of dimension $n$ by boldface Greek letters such as
$\balp = (\alp_1, \dots, \alp_n)$, with $0 \leq \alp_\mu$, 
$\mu=1,\ldots,n$; by an abuse 
of notation, we denote with $\lam$ the multi--index such that 
$\alp_{\mu}=0$, if $\mu\neq \lam$, $\alp_{\mu}= 1$, if 
$\mu=\lam$.
We also set $|\balp| \byd \alp_{1} + \dots + \alp_{n}$.
The charts induced on $J_r\bY$ are denoted by $(x^\lam,y^i_{\balp})$, with $0
\leq |\balp| \leq r$; in particular, we set $y^i_{\bf{0}}
\equiv y^i$. The local bases of vector fields and $1$--forms on $J_r\bY$ induced by
the coordinates above are denoted by $(\der_\lam ,\der^{\balp}_i)$ and $(d^\lam,d^i_{\balp})$,
respectively.

The {\em contact maps\/} on jet spaces \cite{MaMo83a} induce
the natural complementary fibered
morphisms over the affine fiber bundle $J_r\bY \to J_{r-1}\bY$
\bEq\label{affine1}
\cD_{r} : J_r\bY \ucar{\bX} T\bX \to TJ_{r-1}\bY \,,
\qquad 
\vartht_{r} : J_r\bY \ucar{J_{r-1}\bY} TJ_{r-1}\bY \to VJ_{r-1}\bY\,, \qquad r\geq 1\,,
\eEq
with coordinate expressions, for $0 \leq |\balp| \leq r-1$, given by
\bEq\label{affine2}
\cD_{r} &= d^\lam\ten {\cD}_\lam = d^\lam\ten
(\der_\lam + y^j_{\balp+\lam}\der_j^{\balp}) \,, \qquad
\vartht_{r} &= \vartht^j_{\balp}\ten\der_j^{\balp} =
(d^j_{\balp}-y^j_{{\balp}+\lam}d^\lam)
\ten\der_j^{\balp} \,,
\eEq
and the natural fibered splitting \cite{MaMo83a,MaSa00,Sau89}
\bEq
\label{jet connection}
J_r\bY\ucar{J_{r-1}\bY}T^*J_{r-1}\bY =
J_r\bY\ucar{J_{r-1}\bY}\left(T^*\bX \oplus V^{*}J_{r-1}\bY\right)\,.
\eEq

\medskip

Let us consider the following dual exact sequences of vector bundles over $\bY$:

\bEq\label{conn1}
0 \to V\bY \hookrightarrow T\bY \to \bY\ucar{\bX}T\bX \to 0 \,,
\qquad
0 \to \bY\ucar{\bX}T^{*}\bX \hookrightarrow T^{*}\bY \to V^{*}\bY \to 0 \,.
\eEq

\bDf
A {\em connection} on the fiber bundle $\bY \to\bX$ is defined by the dual linear bundle
morphisms over $\bY$
\bEq\label{split1}
\bY\ucar{\bX}T\bX \to T\bY\,, \qquad V^{*}\bY\to T^{*}\bY
\eEq
which split the exact sequences \eqref{conn1}.\END
\eDf

\bRm\label{pull--back}
Let $\Gam^{i}_{\lam}$ be the local components of the connection $\Gam$.
The above linear morphisms over $\bY$ yield uniquely a horizontal tangent--valued $1$--form
on $\bY$, which we denote by $\Gam=d^{\lam}\ten(\der_{\lam}+\Gam^{i}_{\lam}\der_{i})$ and
which projects
over the soldering form on $\bX$.
Dually, a connection $\Gam$ on $\bY$ can be also represented
by the vertical--valued $1$--form $\Gam=(d^{i} -
\Gam^{i}_{\lam}d^{\lam})\ten\der_{i}$ (see \cite{MaSa00}).
Taking this into account, the canonical splitting \eqref{jet connection} provides the horizontal 
splitting \, $J_{1}\bY\ucar{\bY}T\bY\simeq J_{1}\bY\ucar{\bY}(V\bY\udir{\bY} H\bY)$.
\END
\eRm

\bPr\label{affinesec} {\em \cite{GMS97,Sau89}}
There is a one--to--one correspondence between the connections $\Gam$ on a fiber bundle
$\bY\to\bX$ and the global sections 
\,$\Gam:\bY\to J_{1}\bY$\,
of the affine jet bundle $J_{1}\bY\to \bY$.
\ePr

\medskip

In the following a relevant role is played by the composition of fiber bundles
\bEq\label{compositebundle}
\bY\to\Tht \to \bX\,,
\eEq
where $\pi_{\bY\bX}: \bY\to \bX$, $\pi_{\bY\Tht}:\bY\to\Tht$ and $\pi_{\Tht\bX}:
\Tht\to\bX$ are fiber bundles. The above composition was introduced under the name of {\em
composite fiber bundle} in
\cite{GMS97,MaSa98,Sar95} and shown to be useful for physical applications, \eg for the
description of mechanical systems with time--dependent parameters.
We recall some structural properties of composite fiber bundles
\cite{MaSa00}.

\bPr
Given a composite fiber bundle \eqref{compositebundle}, let $h$ be a global section of the
fiber bundle $\pi_{\Tht\bX}$. Then the restriction $\bY_{h}\byd h^{*}\bY$ of the fiber bundle
$\pi_{\bY\Tht}$ to $h(\bX)\sub \Tht$ is a subbundle $i_{h}:\bY_{h}\hookrightarrow\bY$ of the
fiber bundle $\bY\to \bX$.
\ePr

\bPr
Given a section $h$ of the fiber bundle $\pi_{\Tht\bX}$ and a section $s_{\Tht}$ of the fiber
bundle $\pi_{\bY\Tht}$ their composition $s=h\,\com\, s_{\Tht}$ is a section of the composite
bundle
$\bY\to\bX$. Conversely, every section $s$ of the fiber bundle $\bY\to\bX$ is the
composition of the section $h=\pi_{\bY\Tht}\,\com\, s$ of $\pi_{\Tht\bX}$ and some section
$s_{\Tht}$ of $\pi_{\bY\Tht}$ over the closed submanifold $h(\bX)\sub\Tht$.
\ePr

\subsection{Connections on composite bundles}\label{compositeconn}

We shall be concerned here with the description of connections on
composite fiber bundles. We will follow the notation and main results stated in \cite{MaSa00};
see also \cite{CaKo93}.

We shall denote by $J_{1}\Tht$, $J_{1}^{\Tht}\bY$ and $J_{1}\bY$, the jet
manifolds of the fiber bundles $\Tht\to \bX$, $\bY\to\Tht$ and $\bY\to\bX$ respectively.

\medskip

Let $\gam$ be a connection on the composite bundle $\pi_{\bY\bX}$ projectable over a
connection $\Gam$ on $\pi_{\Tht\bX}$, \ie such
$J_{1}\pi_{\bY\Tht}\,\com\,\gam\,=\,\Gam\,\com\,\pi_{\bY\Tht}$.
Let $\mathfrak{H}_{\Tht}$ be a connection on the fiber bundle $\pi_{\bY\Tht}$. 
Given a
connection $\Gam$ on $\pi_{\Tht\bX}$, there exists a canonical morphism over $\bY$
\cite{MaSa00,Sau89},
$\rho: J_{1}\Tht\ucar{\bX}J_{1}^{\Tht}\bY\to J_{1}\bY$, which sends
$(\Gam,\mathfrak{H}_{\Tht})$, into the {\em composite connection} $\gam\byd
\mathfrak{H}_{\Tht}\com \Gam$ on
$\pi_{\bY\bX}$, projectable over $\Gam$.

\bRm\label{restriction}
Let $h$ be a section of $\pi_{\Tht\bX}$. Every connection
$\mathfrak{H}_{\Tht}$ induces the pull--back connection $\mathfrak{H}_{h}$ on the subbundle
$\bY_{h}\to\bX$. We recall that the composite connection
$\gam\,=\,\mathfrak{H}_{\Tht}\,\com\,\Gam$ is reducible to $\mathfrak{H}_{h}$ if and only if
$h$ is an integral section of $\Gam$.\END\eRm

\medskip

We have the following exact sequences of {\em vector bundles
over a composite bundle $\bY$}:
\bEq
0\to V_{\Tht}\bY\hookrightarrow V\bY\to\bY\ucar{\Tht}V\Tht\to 0\,, \qquad 
0\to \bY\ucar{\Tht}V^{*}\Tht \hookrightarrow V^{*}\bY\to V^{*}_{\Tht}\bY \to 0\,,
\eEq
where $V_{\Tht}\bY$ and $V^{*}_{\Tht}\bY$ are the vertical tangent and cotangent bundles to the
bundle
$\pi_{\bY\Tht}$.

\bRm
Every connection $\mathfrak{H}_{\Tht}$ on $\pi_{\bY\Tht}$ provides the dual splittings
\bEq
VY = V_{\Tht}\bY\udir{\bY}\mathfrak{H}_{\Tht}(\bY\ucar{\Tht}V\Tht)\,, \qquad
V^{*}Y = \bY\ucar{\Tht}V^{*}\Tht \udir{\bY}\mathfrak{H}_{\Tht}(V^{*}_{\Tht}\bY)\,,
\eEq
of the above exact sequences.

By means of these splittings we can construct the {\em vertical covariant differential} on the
composite bundle $\pi_{\bY\bX}$, \ie the first order differential operator 
\bEq\label{verdiff}
\Del_{\mathfrak{H}}: J_{1}\bY \to T^{*}\bX
\udir{\bY} V^{*}_{\Tht}\bY\,.
\eEq
The restriction of $\Del_{\mathfrak{H}}$, induced by a section $h$ of $\pi_{\Tht\bX}$, coincides
with the covariant differential on
$\bY_{h}$ relative to the pull--back connection $\mathfrak{H}_{h}$ \cite{MaSa00}.
\eRm\END

%------------------------------------------------------------------------------------------
\section{Hamiltonian formalism for field theory}\label{3}
%------------------------------------------------------------------------------------------

We recall now that the covariant Hamiltonian field theory can be conveniently formulated in terms
of Hamiltonian connections and Hamiltonian forms \cite{GMS97,Sar95}. Here we shall construct a
Hamiltonian formalism for field theory as a theory on the composite bundle $\bY \to\Tht\to \bX$,
with $\pi_{\Tht\bX}: \Tht \to \bX$ a {\em line bundle} having local fibered coordinates $(x^{\lam},
\tau)$.

\medskip

Let us now consider the {\em extended Legendre bundle} $\Pi_{\Tht}\byd
V^{*}\bY\wed (\For^{n-1}T^{*}\Tht)\to \bX$. There exists the canonical isomorphism
\bEq
\Pi_{\Tht}\simeq \For^{n}T^{*}\Tht\uten{\bY}V^{*}\bY\uten{\bY}T\Tht\,.
\eEq

\bDf
We call the fiber bundle $\pi_{\bY\Tht}: \bY\to\Tht$ the {\em abstract event space} of the field
theory. The {\em configuration space} of the field theory is then the first order jet manifold
$J^{\Tht}_{1}\bY$. 

The {\em abstract Legendre bundle} of the field
theory is the fiber bundle
$\Pi_{\Tht}\to\Tht$.
\END\eDf

\medskip

Let now $\mathfrak{H}_\Tht$ be a connection on $\pi_{\bY\Tht}$ and $\Gam_{\Tht}$ be a
connection on
$\pi_{\Tht\bX}$. We have the following non--canonical isomorphism
\bEq\label{iso}
\Pi_{\Tht}\simeq_{(\mathfrak{H}_\Tht,\Gam_{\Tht})}
\For^{n}T^{*}\Tht\uten{\bY}[(\bY\udir{\Tht}V^{*}\Tht)\udir{\bY}
\mathfrak{H}_{\Tht}(V^{*}_{\Tht}\bY)]\uten{\bY}(V\Tht \udir{\Tht}H\Tht)
\,.
\eEq

In this perspective, we consider the canonical bundle monomorphism over $\bY$ providing the
tangent--valued Liouville form on $\Pi_{\Tht}$, \ie
\bEq
\vartht_{\bY}: \Pi_{\Tht}\hookrightarrow \For^{n+1}T^{*}\bY\uten{\bY}(V\Tht \udir{\Tht}
H\Tht)\,,
\eEq
the coordinate expression of which is 
\bEq
\vartht_{\bY}=
p^{\lam}_{i}d^{i}\wed\ome\ten\der_{\lam}\ten\der_{\tau}\simeq
p^{\lam}_{i}\vartht^{i}\wed\ome_{\lam}\ten\der_{\tau}\,,
\eEq
where $\vartht^{i}$ are generators of vertical $1$--forms (\ie contact forms) on $\bY$ and
``$\simeq$'' is the isomorphism defined by \eqref{iso}.

The polysymplectic form $\Ome_{\bY}$ on $\Pi_{\Tht}$ is then intrinsically defined by
\beq
\Ome_{\bY}\rfloor
\psi=d(\vartht_{\bY}\rfloor\psi)\,,
\eeq 
where $\psi$ is an arbitrary $1$--form on $\Tht$; its coordinate expression is given
by 
\bEq\label{multisympl}
\Ome_{\bY}=
dp^{\lam}_{i}\wed d^{i}\wed\ome\ten\der_{\lam}\ten\der_{\tau}\simeq
dp^{\lam}_{i}\wed \vartht^{i}\wed\ome_{\lam}\ten\der_{\tau}\,.
\eEq

\bRm
The polysymplectic form \eqref{multisympl} is related to the kind of {\em `special'
multisymplectic structures} on vector bundles studied from a topological point of view in
\cite{Win01}.
\END\eRm

Let $J_{1}\Pi_{\Tht}$ be the first order jet manifold of the extended Legendre
bundle
$\Pi_{\Tht}\to\bX$. By
Proposition \ref{affinesec} a connection $\gam$ on the extended Legendre bundle is in
one--to--one correspondence with global sections of the affine bundle $J_{1}
\Pi_{\Tht}\to \Pi_{\Tht}$.

\bDf
A connection $\gam$ on the extended Legendre bundle $\Pi_{\Tht}$ is said to be a {\em Hamiltonian
connection} iff the exterior form
$\gam\rfloor\Ome_{\bY}$ is closed.
\END\eDf

As a straightforward application of the relative Poincar\'e lemma we have 
then \cite{MaSa00} the following.

\bPr
Let $\gam$ be a Hamiltonian connection on $\Pi_{\Tht}$ and $\bU$ be an open subset of $\Pi_{\Tht}$.
Locally, we have 
\bEq
\gam\rfloor\Ome_{\bY}=dp^{\lam}_{i}\wed \vartht^{i}\wed\ome_{\lam}\ten\der_{\tau} - d\cH \wed
\ome \byd dH
%\ten \der_{\tau}
\,,
\eEq
where $\cH: \bU\sub\Pi_{\Tht} \to V\Tht$.
\ePr

\bDf
The local mapping $\cH: \bU\sub\Pi_{\Tht} \to V\Tht$ is called a {\em Hamiltonian}. 
The form
$H$ on the extended Legendre bundle $\Pi_{\Tht}$ is called a {\em Hamiltonian form}. 
\END\eDf

Every Hamiltonian form $H$ admits a Hamiltonian connection $\gam_{H}$ such that
\bEq\label{hamconn}
\gam_{H}\rfloor\Ome_{\bY}= dH\,.
\eEq

\medskip

Let now set $\bar{p}^{\lam}_{i}\byd {p}^{\lam}_{i}\der_{\tau}$. Then the Hamiltonian form $H$ is the
Poincar\'e--Cartan form of the {\em  Lagrangian}
$L_{H}=(\bar{p}^{\lam}_{i}y^{i}_{\lam} -\cH)\ome$ on $J_{1} \Pi_{\Tht}$, with values in $V\Tht$.

\bDf
The {\em Hamilton operator} for $H$ is defined as the Euler--Lagrange operator associated with
$L_{H}$, namely:
\bEq\label{Hamop}
\cE_{H}: J_{1} \Pi_{\Tht}\to T^{*}\Pi_{\Tht}\wed\For^{n}T^{*}\bX\,.
\eEq
\END\eDf

The kernel of the Hamilton operator \eqref{Hamop}, \ie the Euler--Lagrange equations for $L_{H}$, is
an affine closed embedded subbundle of $J_{1}\Pi_{\Tht}\to \Pi_{\Tht}$, locally given by the
equations
\bEq\label{hamilton}
y^{i}_{\lam}=\der^{\lam}_{i}\cH\,, \qquad\qquad\bar{p}^{\lam}_{\lam \,i}=-\der_{i}\cH\,.
\eEq

\bDf\label{def1}
The kernel of the Hamilton operator defines the {\em covariant Hamilton 
equations} \eqref{hamilton} on
the extended Legendre bundle $\Pi_{\Tht}\to\bX$.
\END\eDf

\bRm\label{ker}
Notice that a global section of $\textstyle{ker}\cE_{H}\to\Pi_{\Tht}$ is a 
Hamiltonian connection $\gam_{H}$ satisfying relation \eqref{hamconn}.
\END\eRm

In the sequel we state the main result of this note, which points out 
the relation with the standard polysymplectic approach 
(for a review of the  topic see \eg
\cite{Ded77,Go91,Kan98,Kij73,KiTu79,Krup01} and references quoted 
therein). The basic idea is that the present geometric formulation 
can be interpreted as 
a suitable generalization to field theory of the  so--called {\em homogeneous formalism} 
for Mechanics. 

\medskip

Let $\Del_{\Tilde{\mathfrak{H}}}$ be the vertical covariant differential 
(see \eqref{verdiff}) relative to the connection $\Tilde{\mathfrak{H}}_{\Tht}$ on 
the abstract Legendre bundle $\Pi_{\Tht}\to\Tht$.

\bDf\label{def2}
We define the {\em abstract covariant Hamilton equations} to be the 
kernel of the first order differential operator $\Del_{\Tilde{\mathfrak{H}}}$.
\END\eDf

\bLm\label{main}
Let $\gam_{H}$ be a Hamiltonian connection on $\Pi_{\Tht}\to\bX$. 
Let $\Tilde{\mathfrak{H}}_{\Tht}$ and $\Gam$ be connections on 
$\Pi_{\Tht}\to\bY$ 
and $\Tht\to\bX$, respectively. Let $\sig$ and $h$ be sections of the 
bundles $\pi_{\bY\Tht}$ and $\pi_{\Tht\bX}$, respectively.

Then the standard Hamiltonian connection on $\Pi_{\Tht}\to\bX$ turns out to
be the pull--back connection $\Tilde{\mathfrak{H}}_{\phi}$ induced on the subbundle
$\Pi_{\Tht\,\phi}\hookrightarrow \Pi_{\Tht}\to\bX$ by the section $\phi=h\com 
\sig$ of $\bY\to\bX$.
\eLm

\bPf
The abstract Legendre bundle is in fact a composite bundle 
$\Pi_{\Tht}\to\bY\to \Tht$, so that it is possible to apply the 
results concerning connections on composite bundles recalled in Subsection \ref{compositeconn}, 
for any 
connection $\Tilde{\mathfrak{H}}_{\Tht}$. Our claim then 
follows for any section $\phi$ of the composite bundle 
$\bY\to\Tht\to\bX$ of the type $\phi=h\com \sig$, since the extended 
Legendre bundle $\Pi_{\Tht}\to\bX$ can be also seen as the composite 
bundle $\Pi_{\Tht}\to\bY\to\bX$.
\QED\ePf

We can then state our main result as follows.

\bTh\label{final}
Let $\Del_{\Tilde{\mathfrak{H}}\,, \phi}$ be the covariant differential on the subbundle
$\Pi_{\Tht\,\phi}\hookrightarrow \Pi_{\Tht}\to\bX$ relative to the 
pull--back connection $\Tilde{\mathfrak{H}}_{\phi}$. The kernel of 
$\Del_{\Tilde{\mathfrak{H}}\,, \phi}$ coincides with the Hamilton--De Donder 
equations of the standard polysymplectic approach to field theories.
\eTh

\bPf
It is a straightforward consequence of Lemma \ref{main} together with 
Definitions \ref{def1} and \ref{def2}.
\QED\ePf

\bRm
Our approach provides a suitable geometric interpretation of
the canonical theory of gravity and gravitational energy, as presented in \cite{Kij01}, where
$\tau$ plays the role of a {\em parameter} and enables one to consider the gravitational energy
as a `{\em gravitational charge}'. This topic is currently under investigation and it will be
developed in a separate forthcoming paper.
\END\eRm

%------------------------------------------------------------------------------------------
\section{Acknowledgments}
%------------------------------------------------------------------------------------------

One of the authors (M. P.) is grateful to J. Kijowski for helpful comments concerning his
lectures on {\em Canonical Gravity} held in Levo\v ca, August 2000. Thanks are due to R.
Vitolo for useful remarks. M. P. and E. W. also acknowledge the kind invitations at the Department of
Mathematics {\em E. De Giorgi} of the University of Lecce, October 2000 and August--September 2001.
This paper has been written within the GNFM--INdAM research project {\em Formalismo Hamiltoniano
in teoria dei campi} and the University of Torino project {\em Giovani Ricercatori 2001}.

%------------------------------------------------------------------------------------------
% B I B L I O G R A P H Y
%------------------------------------------------------------------------------------------

\end{document}